\documentstyle[nato,epsfig]{crckapb}

\begin{opening}
\title{Quantum Computation and Spin Electronics
}

\author{D.P. DiVINCENZO (1), G. BURKARD (2), D. LOSS (2), E. V. SUKHORUKOV
(2)}
\institute{(1) IBM Research Division, T. J. Watson Research Center\\
           PO Box 218, Yorktown Heights, NY 10598 USA\\
(2) Dept. of Physics and Astronomy, University of Basel,\\
Klingelbergstrasse 82, CH-4056 Basel, Switzerland}

\end{opening}

\runningtitle{Quantum Computation}

\begin{document}

\begin{abstract}
\footnote{Published in {\em Quantum
Mesoscopic Phenomena and Mesoscopic Devices in Microelectronics},
eds. I. O. Kulik and R. Ellialtioglu 
(NATO Advanced Study Institute, Turkey, June 13-25, 1999).}
In this chapter we explore the connection between mesoscopic physics
and quantum computing.  After giving a bibliography providing a
general introduction to the subject of quantum information processing,
we review the various approaches that are being considered for the
experimental implementation of quantum computing and quantum
communication in atomic physics, quantum optics, nuclear magnetic
resonance, superconductivity, and, especially, normal-electron solid
state physics.  We discuss five criteria for the realization of a
quantum computer and consider the implications that these criteria
have for quantum computation using the spin states of single-electron
quantum dots.  Finally, we consider the transport of quantum
information via the motion of individual electrons in mesoscopic
structures; specific transport and noise measurements in coupled
quantum dot geometries for detecting and characterizing electron-state
entanglement are analyzed.
\end{abstract}

\section{Brief Survey of the History of Quantum Computing}

The story of why quantum computing and quantum communication are
theoretically interesting and important has been told in innumerable
places before, and we will just point the reader to some of those.
The ``prehistory'' of quantum computing (up to 1994) consists of the
tinkerings of a small number of visionaries on the question of how
data could be processed if bits could be put into quantum
superpositions of states.  The idea of a quantum gate was introduced,
the basic possibilities of quantum algorithms were set forth, quantum
communication (in the form of quantum cryptography) was well
developed, and some rudimentary ideas of how physical systems could be
made to implement quantum computing were considered.  Actually, it is
humbling to reflect on how few people it took (no more than about 15,
we would say) to launch a set of basic concepts which have now come to
occupy the attention of some hundreds of researchers.

Here is a bibliography of recent and not-so-recent review articles
which would bring the reader up to speed on quantum computation and
related areas.  Those interested in seeing how complete the
``prehistoric'' perspective was should consult Ekert's influential
paper at Atomic Physics 14 \cite{Ekert}, which helped to make atomic
physicists the earliest participants in the attempt to build a quantum
computer.  Another early but influential review which is a very good
source for the Shor algorithm is \cite{EJ}.  An early general overview
is \cite{Ben}, and an early simple discussion of quantum gate
constructions is \cite{DivITP}.  More recent general overviews are
\cite{Bar} and \cite{Steane}.  \cite{BS} and \cite{MMM} give a set of
ideas for the future direction of quantum communication and
information theory.  Many surveys have considered the question of how
quantum computers might be implemented in actual physical devices; for
the solid-state type of implementations which we will discuss below,
we might mention \cite{Scietc} besides some of the other articles
already cited.

\section{Creating the Quantum Computer}

Of course, the reason for this contribution to appear in a school on
mesoscopic quantum phenomena is that we hope, ultimately, that out of
mesoscopic physics will emerge the capability actually to build a
working, scalable quantum computer.  Microelectronic mesoscopic
devices will inevitably emerge, of course, but whether they will be
usable for quantum computation will depend on whether they manage to
satisfy a very specific set of requirements.

\subsection{Five Criteria for Building a Quantum Computer}

These requirements can be boiled down to a list of five criteria.  We
have written frequently about these five criteria
before\cite{MMM,Scietc}, but they are worth considering in detail here
again, since we have found that these criteria provide a unifying
framework for these investigations which encompass an astonishingly
broad and rich range of fundamental physics.  Further investigations
have endowed this simple list of five with more and more richness and
interest, and show how exciting the building of a quantum computer
will be from the point of view of novel and basic physics.

Before getting into the serious work, we find it amusing to
state the criteria stripped of all physics language and posed
purely as a set of requirements for building a computer.  In
this light our five points are extremely trivial:
\begin{enumerate}
\item The machine should have a collection of bits.
\item It should be possible to set all the memory bits to 0 before the
start of each computation.
\item The error rate should be sufficiently low.
\item It must be possible to perform elementary logic operations
between pairs of bits.
\item Reliable output of the final result should be possible.
\end{enumerate}
It has to be admitted that no great brainwork is required to
arrive at this set of basic attributes which a computer ought to
possess.  But let us take the next step and translate these
into a physics language, to specify what it really takes to
achieve these very basic and simple requirements in a quantum
setting:
\begin{enumerate}
\item A physical system with a collection of well characterized
quantum two-level systems (qubits) is needed.  Each qubit should be
separately identifiable and externally addressable.  The dynamics of
the system should be under sufficient control that these qubits are
never excited into any third level.  It should be possible to add
qubits at will.
\item It should be possible to, with high accuracy, completely
decouple the qubits from one another, and it should be possible to
start an experiment by placing each qubit in its lower (0) state.
\item The decoherence time of these qubits should be long, ultimately
up to $10^4$ times longer than the ``clock time'' (see the next
requirement).
\item Logic operations should be doable.  This involves having the
one-body Hamiltonian of each qubit under independent and precise
control (this gives the one-bit gates).  Two-body Hamiltonians
involving nearby qubits should also be capable of being turned on
and off under external control (these are the two-bit gates).  In a
typical operation, the one- or two-body Hamiltonian will be turned on
smoothly from zero to some value and then turned off again, all within
one clock cycle; the integral of this pulse should be
controllable to again about one part in $10^4$.
\item Projective quantum measurements on the qubits must be doable.
It is useful, but not absolutely necessary, for these measurements to
be doable fast, within a few clock cycles.  It is also useful, but not
necessary, for the measurement to have high quantum efficiency (say
50\%).  If the quantum efficiency is many orders of magnitude lower,
then the quantum computation must be done in an ``ensemble'' style, in
which many identical quantum computers are running simultaneously.  As
with all the other requirements, these measurements absolutely must be
qubit specific.
\end{enumerate}
The reader will notice that there is a lot more to be said about these
criteria as physics requirements than as computer requirements.
Indeed, we will see in the examples we review below that these
criteria involve not just innovations in materials preparation (for
numbers 1 and 3), not just novel device design and fabrication (for
number 4), not just new, precision high-speed electronics at the
nanometer scale (number 4), not just unprecedented capabilities for
ultrasensitive metrology (number 5), but all of these at once!

\subsection{Potential Realizations}

There is a remarkably long list of physical systems that have been
proposed, and are under active experimental investigation, for the
creation of a quantum computer.  This list is remarkable not only for
its length but also for its diversity; it seems that virtually every
area of quantum physics has a candidate.  (This is not quite true: we
know of no proposed quantum computer emerging from high-energy
particle physics.)

Since non-solid state implementations are not the focus of this
article, we will merely mention the many approaches in this category
and give a bibliography: In atomic physics, the original proposal by
Cirac and Zoller\cite{CZ} considered the internal states of ions as
the qubits, coupled by the common vibrational mode of the ions in the
trap.  Later this group\cite{Pel} introduced a variant in which the
coupling is provided by a common cavity electromagnetic mode.  This
work has inspired several analogous solid state proposals which we
will mention below.  Many other variants of the cavity quantum
electrodynamics schemes have been proposed; in some of these the qubit
could be transmitted from one subprocessor to another as a single
photon (in an optical fiber, say)\cite{vanE}; these schemes are
intriguing in that they offer an idea of how to do distributed quantum
computing and distributed decision problems.

A very recent development from atomic physics, neutral-atom optical
lattices, has been proposed to provide the necessary tools to build a
quantum computer\cite{Deuts}.  Here, the ability to move trapped atoms in a
state-dependent way by changing the (classical) phase of the trapping
laser beams is exploited to obtain the necessary conditional dynamics
to do two-bit gates.  There are many ideas for mixing and matching
these various atomic-physics proposals.

Another large area of research has involved bulk NMR of small
molecules containing sets of spin-1/2 nuclei in
solution\cite{Chu,Cor}.  This represents an ``ensemble'' approach,
exploiting the ability to read out the result of quantum computation
with only low quantum efficiency measurements (criterion 5 above).
Unfortunately these schemes presently do not satisfy criterion 2
above, but in other respects it represents a possible approach
to a scalable quantum computer.

An ``almost'' solid state approach, which we will mention just because
it illustrates that quantum computing proposals are coming from just
about every area of physics these days, is one proposed by Platzman
and coworkers\cite{Pla} involving two dimensional crystalline layers
of electrons that can be trapped near the surface of liquid He.  We
are not knowledgeable enough to comment on this proposal in any detail
(it seems to involve elements of the quantum dot proposals we will
discuss in a moment, plus some ideas from the atomic physics
proposals), but it is interesting to see contact made with another,
evidently quite well developed, area of research in quantum physics.

\section{Solid State Proposals}

There have been almost as many proposals for solid state
implementations of quantum computers as all the other proposals put
together.  We think there are some clear reasons for this: we believe
that solid state physics is the most versatile branch of physics, in
that almost any phenomenon possible in physics can be embodied in a
correctly designed condensed matter system.  Even rather esoteric
properties of field theories which are of interest in high-energy
physics have useful realizations, for example, in the fractional
quantum Hall effect.  A related reason is that solid state physics,
being so closely allied with computer technology, has exhibited great
versatility over the years in the creation of artificial structures
and devices which are tailored to show a great variety of desired
physical effects.  This has been exploited very powerfully to produce
ever more capable computational devices.  It would be natural to
extrapolate to say that this versatility will extend to the creation
of solid state quantum computers as well; the plethora of proposals
would indicate that this is indeed true, although time will tell
whether any of these proposals will actually provide a successful
route to a quantum computer!

\subsection{The Spintronic Quantum Dot Proposal}

We will first go through, in some detail, the proposals that
we\cite{LD} have made for the use of coupled quantum dot arrays for
quantum computation; we will also discuss several closely related
proposals such as the ``Kane model''\cite{Kane,Vrijen}.  Our proposal
may rightly be termed a ``spintronic'' proposal, as it attempts to use
as much as possible the tools of standard electronics (control via
voltage gates, readout by current detection) to accomplish quantum
computing; but our proposal departs from conventional electronics in
that it uses the {\em spin} of the electrons as the basic information
carrier.

We think that {\bf Criterion 1} above is most naturally satisfied by a
genuine spin-1/2 system, which by its nature has a doubly-degenerate
ground state to serve as a qubit.  As we will discuss, the prospects
for long coherence times are better for this qubit than for many other
``pseudo-spin'' degrees of freedom that might be chosen.  (But we will
mention several other possible ones being considered at the end of
this section.)

\begin{figure}
\centerline{\psfig{file=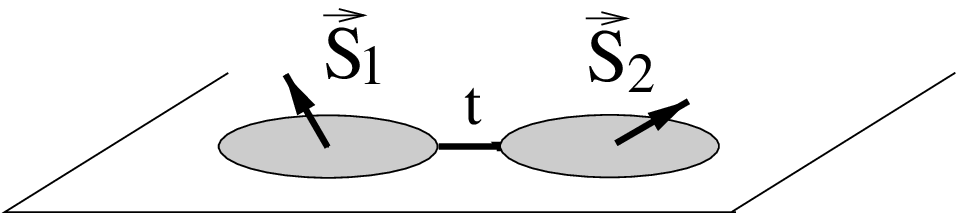,width=8cm}}
\centerline{\psfig{file=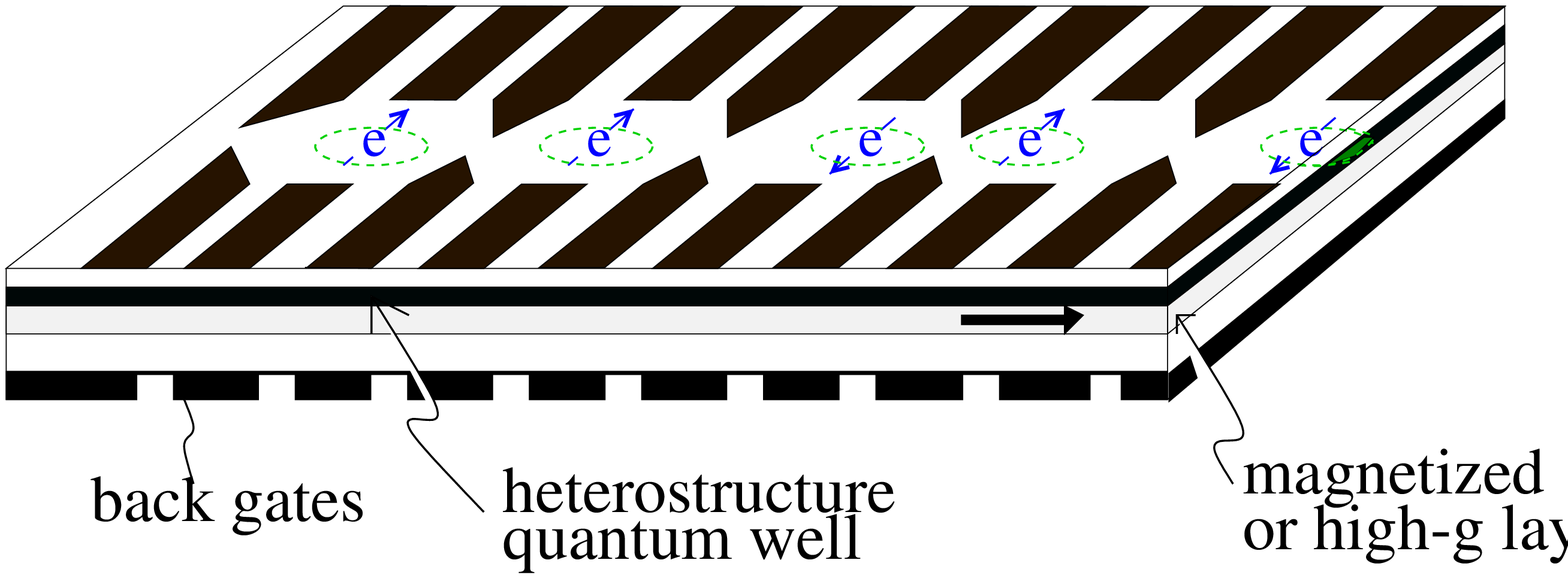,width=10cm}}
\caption{\label{dots} Above: Two coupled quantum dots, between which
electrons can tunnel with amplitude $t$. This tunneling leads to an
effective spin interaction $J\sim t^2/U$ (where $U$ is the on-site
Coulomb repulsion) between the excess spins ${\bf S}_1$ and ${\bf
S}_2$ in the dots, which can be controlled by a number of external
parameters.  In principle, any type of tunnel-coupled confined
structure is a candidate for the ``spintronics'' quantum computing
proposal. E.g., the dots can be defined electrically in a
two-dimensional electron gas (as suggested by this drawing), or they
can be vertically coupled dots, or even atoms, {\em etc}.  Below:
Concept for quantum-dot array device.}
\end{figure}

We have to specify how one obtains the individually identifiable and
addressable qubits of Criterion 1.  Our scheme for doing this involves
an extension of well-demonstrated techniques in the area of mesoscopic
quantum-dot physics.  In particular, we are interested in quantum dots
created by lateral confinement in a two-dimensional electron gas.  The
structure is illustrated in Fig.~\ref{dots}.  To have qubits in this
structure (one per quantum dot) using the electron spin, it is
necessary for the electron number to be controlled precisely; this is
accomplished using the well-documented Coulomb blockade effect.  The
electron number must be fixed at an odd value in order for the spin
quantum number to be 1/2.  This can be achieved if the electron number
can be controllably tuned down to one (technically difficult in
structures in which the Coulomb blockade is probed by transport
measurements, quite feasible in structures where the Coulomb blockade
is sensed capacitively); alternatively, a consensus seems to be
developing that a dot with a spin quantum number of 1/2 gives a
characteristic zero-bias anomaly in transport which signals a Kondo
effect.  This approach will have to be used with caution, however--the
spin quantum number of a dot may readily switch from 1/2 to 3/2 as a
function of the dot shape\cite{Brouwer}; it is of course essential
that this switching does not occur between the configuration in which
a qubit is characterized and that in which it is used.

Criterion 1 also requires that the next higher lying state should
never (or hardly ever) be occupied; this requires that dots be
sufficiently small that the next level is inaccessible to thermal
excitation or non-adiabatic excitation.  It appears that this
requirement will be quite readily satisfied for dots of the 10s of
nanometer sizes that are already studied, at normal (l-He) cryogenic
temperatures.  As for extendibility, as our Figure 1 suggests this is
``simply'' a matter of making an array of dots, either one or two
dimensional.  Of course, this cartoon does not do justice to the
actual complexity of the multilevel wiring layout required to address
a large array of qubits; we hope we are justified in considering this
an ``engineering'' matter, although by no means a trivial one.

In the other spin-qubit scheme we have considered\cite{Vrijen}, the
``quantum dot'' is provided by a single donor impurity atom lying near
the surface of a Si-Ge heterostructure.  Such a system is a kind of
natural one-electron quantum dot; the hydrogenic impurity potential
can normally only bind a single conduction-band electron (sometimes it
can bind two: see Criterion 5 below).  The characteristics of this
qubit (its g-factor, orbital size, and so forth) are more nearly
unique than in the lateral quantum dot proposal; this could ultimately
be either an a advantage or a disadvantage.  Also, placing individual
impurities at specified locations near the surface of a semiconductor
is a very daunting technical challenge, although perhaps one that the
technological world will be needing to face in any case in the coming
nanoelectronic age.  In general, Criterion 1 poses very serious
challenges in the art of material preparation and device fabrication,
as our discussion should make clear.

{\bf Criterion 2} is a relatively straightforward one, but involves
additional device-fabrication considerations.  In order for the
quantum-dot or donor-impurity qubit to start in a stable all-0 state,
it is sufficient to place the spins in a several-Tesla magnetic field
at liquid-He temperatures; then the probability of occupying the
all-down ground state is sufficiently high.  In order for this state
to be stable upon removal of the magnetic field, we require a probably
more difficult requirement: the qubits must be, to rather high
accuracy (to be discussed further in the next two criteria), decoupled
from one another.  The device geometry is chosen so that, in the
``resting'' state, the ``point contact'' voltage probes (the pairs of
electrodes between neighboring quantum dots) can be set to a high
repulsive voltage so that the overlap between neighboring quantum-dot
orbitals is negligible.  Turning this voltage to low can increase this
overlap by orders of magnitude, and this is an essential computational
step (see Criterion 4).

In the donor-impurity scheme, top electrodes control the near-surface
band bending of the semiconductor; in the flat-band condition, the
donor hydrogenic orbitals are compact enough that the overlap between
neighbors is negligible.  By suitable band bending, the electron
orbitals can be made much more delocalized, leading to orders of
magnitude more spin-spin interaction.

{\bf Criterion 3} probably involves the must fundamental quantum
physics, in that it relates to the issues of the decoherence of
quantum systems and the transition between quantum and classical
behavior.  Of course, a lot of attention has been devoted in
fundamental mesoscopics research to characterizing and understanding
the decoherence of electrons in small structures.  We remind the
reader, however, that most of what has been probed (say in weak
localization studies or the Aharonov-Bohm effect) is the {\em orbital}
coherence of electron states, that is, the preservation of the
relative phase of superpositions of spatial states of the electron
(e.g., in the upper or lower arm of an Aharonov-Bohm ring).  The
coherence times seen in these investigations are almost completely
irrelevant to the {\em spin} coherence times which are important in
our quantum computer proposal.  There is some relation between the two
if there are strong spin orbit effects, but our intention is that
conditions and materials should be chosen such that these effects are
weak.

Under these circumstances the spin coherence times (the time over
which the phase of a superposition of spin-up and spin-down states is
preserved) can be completely different from the charge coherence
times, and in fact it is known that they can be orders of magnitude
longer.  This was actually one of our prime motivations for proposing
spin rather than charge as the qubit in these structures.  The
experimental measurement of this kind of coherence is not so familiar
in mesoscopic physics, but fortunately it is very familiar in the area
of spectroscopy.  The measurement to probe spin coherence is
essentially equivalent of the characterization of the so-called $T_2$
time in spin resonance.

In fact, we suggest that the quantification of spin coherence in the
quantum dot quantum computer will follow a line very familiar from the
many recent experiments of Awschalom, Kikkawa, and collaborators
\cite{Kikkawa} that have observed spin precession in a variety of
semiconductor materials.  We expect that these experiments will set up
a large array of identical quantum dots; the dots don't have to be
fully ``wired up'' for quantum computation, but they should otherwise
have the same device structure as in the quantum computer (because we
expect that the coherence times should be very device- and
structure-specific).  The spins will be set in a superposition of up
and down in a magnetic field, and the decay of free induction as these
spins precess in the field will be observed.  Conventional ``spin
echo'' tricks should be used to eliminate the effects of residual
inhomogeneities.  Note that this measurement only requires low quantum
efficiency, so it is not nearly so difficult as the quantum
measurements which we will discuss in Criterion 5.

Awschalom and coworkers have already done many such measurements on a
variety of semiconductor systems, and sees decoherence times up to
hundreds of nanoseconds in some structures\cite{Kikkawa}.  For donor
impurities in silicon, traditional electron spin resonance
measurements have seen $T_2$ times for the P donor spin up to hundreds
of microseconds.  As we see decoherence times in the range from
microseconds to milliseconds as acceptable for quantum computation,
these results are very encouraging; but they are not conclusive.
Decoherence times, depending on the details of the quantum degrees of
freedom in the environment that the qubit can become entangled with,
are expected to be very sensitive to the details of the makeup of the
physical device.  Since none of the experiments have been done on an
actual quantum computing structure as we envision it, the existing
results cannot be viewed as conclusive.  Because of this sensitivity
to details, theory can only give general guidance about the mechanisms
and dependencies to be looked for, but cannot make reliable {\em a
priori} predictions of the decoherence times.

In fact there are further complications in store: we know
theoretically that decoherence is not actually fully characterized by
a single rate; in fact, a whole set of numbers is needed to fully
characterize the decoherence process (12 in principle for individual
qubits), and no experiment has been set up yet to completely measure
this space of parameters, although the theory of these measurements is
available.  Even worse, decoherence effects will in principle be
modified by the act of performing quantum computation (during gate
operation, decoherence is occurring in a coupled qubit
system\cite{LD}).  We believe that the full characterization of
decoherence will involve ongoing iteration between theory and
experiment, and will probably be inseparable from the act of building
a reliable quantum computer.

Finally, the decoherence time $\tau_\phi$ by itself is not a figure
of merit of a quantum computer proposal --- the amount of coherent
computation which can be performed depends on the ratio
$\tau_\phi/\tau_s$ where $\tau_s$ denotes the switching
time ($\tau_s^{-1}$ is the ``clock frequency'' of the quantum
computer).

{\bf Criterion 4} also requires an extensive discussion of the physics
of the proposed qubit device.  It is necessary to identify simple,
reliable mechanisms by which specified qubits can be subjected to
one-body (one-bit gates) and two-body (two-bit gates) Hamiltonians
which can be turned on and off in time.  Almost all the complexity of
the proposed devices arises from the need to achieve these capabilities.

In the quantum dot array structure, the two-bit gates are obtained by
a controlled lowering of the potential barrier produced by the ``point
contact'' gates between neighboring quantum dots.  When this barrier
is lowered, the two electrons are brought together, forming,
temporarily, an artificial hydrogen molecule.  The effective spin-spin
interaction that this produces should, to high accuracy, be given by a
Heisenberg interaction $J{\vec S}_i\cdot{\vec S}_{i+1}$, where the
exchange coupling should be tunable up to about 0.1 meV
\cite{Burkard}, with the ``off'' value being many orders of magnitude
lower than this.  The exponentially strong suppression of the coupling
$J$ in the ``off'' state of the gate is essential, because it assures
that no correlated errors occur due to the switching mechanism.
Corrections to the Heisenberg form should arise only from relativistic
effects (spin-orbit coupling); although spin-orbit corrections to the
band parameters like the g-factor are fairly large in GaAs, the
residual relativistic effects on the low-angular momentum
effective-mass conduction band states can be estimated to be
negligibly small \cite{Scietc}(c).

For laterally tunnel-coupled quantum dots in a two-dimensional
electron system it was found that besides electrical gating, an
external magnetic field can be used to switch on and off the spin-spin
interaction \cite{Burkard}. Recently, there has been great interest in
vertically tunnel-coupled dot structures, both in etched vertical
columnar heterostructures\cite{Austing} and in double-layer
self-assembled quantum-dot structures\cite{Luyken}. We have analyzed
the spin-interaction in such vertically coupled quantum
dots\cite{Seelig}, and have found, in addition to the ones known from
laterally coupled dots, a new mechanism which allows the external
control of the spin interaction. In this scenario, two coupled quantum
dots of different size are subject to an external electric field.  The
field shifts the big dot by a larger distance than the small dot,
therefore leading to an effective increase in the inter-dot distance,
which causes an exponential suppression of the spin exchange coupling.

The proposed two-bit gate mechanism in the donor-impurity quantum
computer is conceptually almost identical to the proposals above.  By
weakening the binding of the electron to the impurity by band bending,
the orbitals can be made to spread out so that the overlap of
neighboring orbitals becomes appreciable; the exchange physics, and
the expected effective Hamiltonian, is the same as the quantum dot
case.

It is known that universal, fault tolerant quantum computation can be
obtained with the logic gates obtainable from this nearest neighbor
exchange interaction \cite{Gott}.  For dots of tens of nanometer size,
a smooth (i.e. adiabatic) turning on and off the Hamiltonian on a time
scale of 10s or 100s of picoseconds would be desirable.  (The
smoothness is required so that higher-lying states of the dot are not
unintentionally excited \cite{Burkard}.)  This is technically
feasible, although it requires very high-bandwidth control (10-100
GHz) of the voltages on each individual electrode (Fig.~\ref{dots}) of
the structure, which is a daunting job of microwave engineering.

Especially daunting is the theoretical precision requirement, which is
that the integrated strength of the exchange Hamiltonian should be
controlled to about one part in $10^4$.  This number, based on the
analysis of the efficacy of error correction techniques in quantum
computation, may come down as better error correction schemes are
devised.  Also, if this number were ``only'' $10^2$ in an experiment,
many interesting studies of quantum computation could still be
performed.  As an example, we have described a minimal experimental
test for quantum error correction involving (at least) three coupled
quantum dots\cite{BLDS}.  Time will tell what the ultimate technical
requirements will be.  A fortunate fact about GaAs quantum dot
structures, holding out the hope that precision manipulation will be
possible, is the observation that, when treated correctly, quantum dot
structures show essentially no ``charge switching'' effects which are
the origin of traditional 1/f noise in transport\cite{KM}.  Note that
uncontrolled charge switching is not nearly so great a problem for
spin qubits as for charge qubits, since this switching does not couple
directly to the spin degree of freedom.  But, since the second order
effects of charge motion could change the strength of the exchange
coupling $J$ by more than a part in $10^{-4}$, the ability to suppress
1/f effects will be very important for switching in quantum
computation.

One-qubit gates must also be considered, and these involve rather
different physics.  Theoretically the requirement is very simple for a
spin-1/2 qubit: it must be possible to subject a specified qubit to a
(real or effective) magnetic field of specified direction and
strength.  We have offered many suggestions previously\cite{LD,Burkard} on
how this requirement may be met: by the application of real, localized
magnetic fields using a scanned magnetic particle or nanoscale electric
currents; by the use of a magnetized dot or magnetized barrier
material that the electron can be inserted in and out of by electric
gating; by the judicious choice of g-factor-modulated
materials\cite{MMM}.  We have performed some detailed analysis of this
last mechanism recently, and it is also the preferred mechanism in the
Si-Ge heterostructure scheme (Si and Ge have very different
g-factors), so we will discuss this recent work here.

Due to spin-orbit coupling, the Land\'e g-factor in bulk semiconductor
materials differs from the free-electron value $g_0=2.0023$ and ranges
from large negative to large positive numbers for various
materials. In confined structures such as quantum wells, wires, and
dots, the g-factor is modified with respect to the bulk material and
sensitive to an external bias voltage\cite{Ivchenko}. Here, we study
the simpler case of a layered structure in which the effective
g-factor of electrons is varied by electrically shifting their
equilibrium position from one layer (with g-factor $g_1$) to another
(with another g-factor $g_2\neq g_1$). For simplicity, we use the bulk
g-factors of the layer materials, an approximation which becomes
increasingly inaccurate as the layers become thinner\cite{Kiselev}.

We consider a quantum well (e.g. AlGaAs-GaAs-AlGaAs), in which some
fraction $y$ of the Ga atoms are replaced by In atoms in the upper
half of the heterostructure (we have used $y=0.1$).  The sequence of
layers in the heterostructure is then
Ga$_{1-x}$Al$_{x}$As-GaAs-Ga$_{1-y}$In$_{y}$As-Ga$_{1-x-y}$Al$_{x}$In$_{y}$A
s,
where $x$ denotes the Al content in the barriers (typically around
30\%). Changing the vertical position of the electrons in the quantum
well via top or back gates permits control of the effective g-factor
for the corresponding electrons: If the electron is mostly in a pure
GaAs environment, then its effective g-factor will be around the GaAs
bulk value (g$_{\rm GaAs}$=-0.44) whereas if the electron is in the
InGaAs region, the g-factor will be somewhere between the GaAs and the
InAs values (g$_{\rm InAs}$=-15). We have analyzed the problem of a
single electron in such a structure, neglecting screening due to
surrounding electrons.
This procedure is justified, since we are interested in isolated
electrons located in quantum dots. In a quantum well with a high
electron density, however, many-body effects should be taken
into account.

\begin{figure}
\centerline{\psfig{file=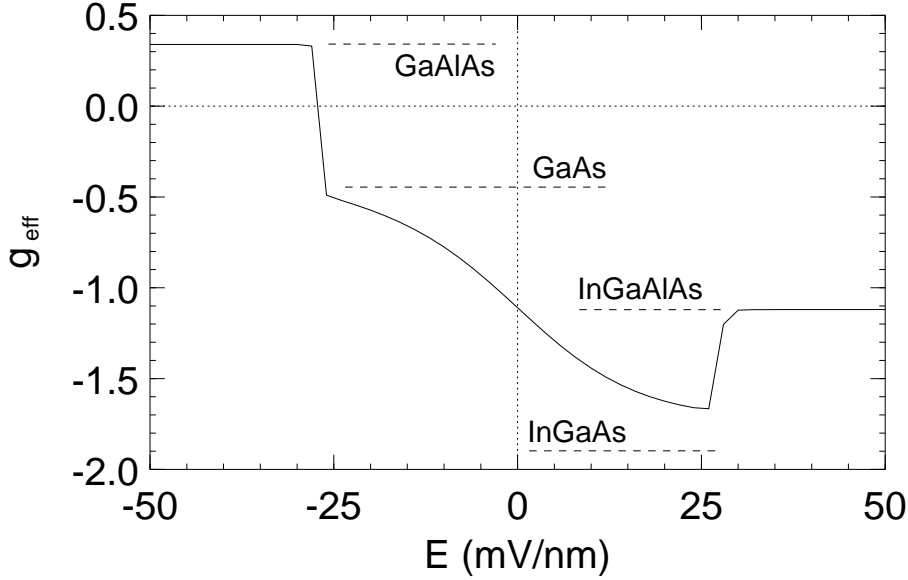,width=12cm}}
\caption{The effective g-factor $g_{\rm eff}$ of electrons confined in
a Ga$_{1-x}$Al$_{x}$As - GaAs - Ga$_{1-y}$In$_{y}$As -
Ga$_{1-x-y}$Al$_{x}$In$_{y}$As heterostructure ($x=0.3$, $y=0.1$) as a
function of the applied electric field $E$ in growth direction.  The
widths of the quantum well and the barriers are $w=w_{\rm B}=10\,{\rm
nm}$.  The g-factors which are used for the materials are indicated
with horizontal lines.}
\label{gfig}
\end{figure}

We have solved the one-dimensional problem,
\begin{equation}
\left[ -\frac{d}{dz}\frac{\hbar^2}{2m(z)}\frac{d}{dz}+V(z)
\right]\Psi(z) = E\Psi(z),
\end{equation}
of a single electron with a spatially varying effective mass $m(z)$
numerically by discretizing it in real space and subsequently performing
exact diagonalization. The potential $V(z)$ describes the quantum
well (conduction band offset $\Delta E_c=270\,{\rm meV}$) and the
electric field $E$ in growth direction. For the effective masses and
g-factors of the various layers we have linearly interpolated between the
GaAs, AlAs, and InAs values. The resulting effective g-factor was
calculated by averaging the g-factor over the electronic
ground-state wavefunction,
\begin{equation}
g_{\rm eff} = \int dz g(z) |\Psi (z)|^2.
\end{equation}
In Fig.~\ref{gfig}, we plot the effective g-factor $g_{\rm eff}$
versus the electric field for a quantum well which is $w=10\,{\rm nm}$
wide. For the barrier thickness we have assumed $w_{\rm B}=10\,{\rm
nm}$. At moderate electric fields, $g_{\rm eff}$ interpolates roughly
between the GaAs and Ga$_{1-y}$In$_{y}$As g-factors. If the electric
energy $eEw_{\rm B}=eU_{\rm B}$ becomes larger than the barrier
$\Delta E_c$, we observe a vertical deconfinement of the electrons. In
our plot (Fig.~\ref{gfig}) the electric deconfinement is clearly seen
as a jump of the effective g-factor to the barrier material value at
$E=\pm \Delta E_c/ew_{\rm B}=\pm 27\,{\rm mV/nm}$.  The electric field
required for a substantial change in $g_{\rm eff}$ is of the order of
$10\,{\rm mV/nm}$, corresponding to a voltage of $100\,{\rm mV}$,
which is about one order of magnitude smaller than the band gap
($1.5\,{\rm eV}$ for GaAs at $T=0$).

Since the g-factor is spatially varying, the Zeeman
coupling influences the electron wavefunction, which in principle
could lead to a non-linear spin splitting $\Delta E(B)$. For the
above materials and parameters however, we find numerically that
the splitting is almost exactly linear,
$\Delta E(B)\simeq g_{\rm eff}\mu_B B$.
The irrelevance of the Zeeman coupling for the orbital wavefunction
is due to the fact that the typical electronic kinetic energy is at
least two orders of magnitude larger than the typical Zeeman energy.

The described quantum well can host an array of electrostatically
defined quantum dots, containing a single excess electron (and thus a
single spin 1/2) each. In order to carry out a single-qubit operation
on one of the spins, the whole system is placed into a homogeneous
magnetic field. By changing the voltage at the electric gate on top of
a single quantum dot, the effective g-factor $g_{\rm eff}$ for the
spin in this quantum dot can be changed by about $\Delta g_{\rm
eff}\approx 1$ with respect to the g-factor of all remaining
spins. This leads to a relative rotation about the direction of ${\bf
B}$ by an angle of roughly $\phi=\Delta g_{\rm eff}\mu_B
B\tau/2\hbar$. The typical switching time $\tau$ for a $\phi=\pi/2$
rotation using a field of $1\,{\rm T}$ is then approximately
$\tau\approx 2\phi\hbar/\Delta g_{\rm eff}\mu_B B\approx 30\,{\rm
ps}$.  Controlling the top gate at $\tau^{-1}\approx 30\,{\rm GHz}$
seems very challenging; we emphasize however that the single-qubit
operation can be done much more slowly (a lower limit is provided by
the spin dephasing time). The switching can be slowed down either by
choosing a smaller $\Delta g_{\rm eff}$ or by replacing $\phi$ by
$\phi+2\pi n$ where $n$ is an integer.

Actually, we should note that there is a substantial degree of
flexibility in how universal quantum computation is achieved.  We have
noted in our original work\cite{LD} that switchable effective magnetic
fields on the dots are not needed for the implementation of one bit
gates, if there are some dots which have a higher static magnetic field,
either because they are magnetized or because of the presence of a
fixed magnetic field gradient.  Then, one-qubit operations can be
effected by swapping the qubit onto the magnetized dot, then swapping
if off again once the desired interaction with the magnetic field
has occurred.

In yet another variant along these lines, Bacon {\em et al.}
\cite{Bac} have very recently shown that, at the price of increasing
the number of quantum-dot spins required to represent each qubit, all
computation can be done with exchange interactions alone, without the
need for any local magnetic fields, except during the measurement
operation.  These workers define a logical qubit as the two-level
system of the singlet states of four spins, in which
\begin{equation}
\begin{array}{lll}
|0_L\rangle&=&|S\rangle\otimes|S\rangle,\\
|1_L\rangle&=&{1\over\sqrt{3}}[|T_+\rangle\otimes|T_-\rangle
-|T_0\rangle\otimes|T_0\rangle+|T_-\rangle\otimes|T_+\rangle].
\end{array}
\end{equation}
Here $|S\rangle$ is the singlet state of two spins and
$|T_{+,-,0}\rangle$ are the three triplet states of two spins.  The
initial preparation of $|0_L\rangle$ is easy; introduce a strong
exchange interaction between pairs of spins (e.g., on the ``even''
bonds but not on the ``odd'' bonds of a one-dimensional chain); the
ground state of this Hamiltonian is the desired state.
Ref. \cite{Bac} shows that all necessary one- and two-qubit gates
between these logical qubits can be done by sequences of exchange
interactions only.  (More work should be done to make these sequences
explicit and short.)  This is possible because the exchange
interactions can move the quantum state vector anywhere within the
total singlet (total spin=0) sector.  Ref. \cite{Bac} also notes that
the measurement of the logical qubit (anticipating Criterion 5 below)
can be done if the first two spins in the four-spin block are measured
in the $z$ bases and the next two are measured in the $x$ basis.

\begin{figure}
\centerline{\psfig{file=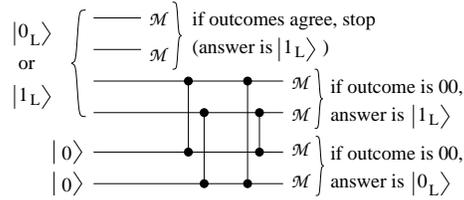,width=6cm}}
\caption{Two ancilla spins set to the $|0\rangle$ state suffice to
measure the logical qubit $|0_L,1_L\rangle$.  First, the first two
spins of the logical qubit are measured in the $z$ basis.  If the
two measurement outcomes agree, then the procedure is finished,
the outcome is $|1_L\rangle$.  Otherwise, the ancilla must be used;
first, the four two-bit gates are performed in the sequence indicated.
The gate is the square-root-of-swap; see \protect\cite{LD,BLDS}.  After
this circuit, all four remaining spins are measured in the $z$ basis.
If the top two measurements give 00, the outcome is $|1_L\rangle$;
if the bottom two measurements give 00, the outcome is $|0_L\rangle$.
No other measurement results are possible.}
\label{figmeas}
\end{figure}

We have recently noted that even this potentially inconvenient
requirement for two different measurement bases can be eliminated.
Only $z$ measurements are needed, if it is assumed that other spins,
initialized in the $|0\rangle$ state ({\em not} $|0_L\rangle$), are
also available, which is possible if some spins are initially cooled
in a uniform external field without being exchange-coupled to their
neighbors.  Then, the measurement protocol of Fig. \ref{figmeas}
suffices.

This little digression about coded qubits illustrates a more general
theme here: the theory indicates that many tradeoffs are possible, in
this case between the number of spins needed and the complexity of the
required gate operations.  Each specific experimental case needs to be
carefully assessed to see what tradeoffs are possible, and how they
can be optimized.

We think that {\bf Criterion 5} hold the prospect for some early
successes in this long road we have laid out to achieving a quantum
computer, successes that will advance some fundamentally interesting
agendas in solid state physics.  This criterion can be simply posed:
how do we measure the state of the spin of a single electron in a
solid?  Actually there is a lot of flexibility in this requirement, in
that high quantum efficiency is not in principle needed: if the
quantum efficiency is 1\%, say, then the computer can still give a
reliable output if the qubit to be read is first copied, say, 200
times, then each of these qubits is subjected to this low-efficiency
measurement. (Note that this does not violate the no-cloning theorem,
because the extra qubits act as ``copies'' only in the basis specified
for the measurement).  But there are various reasons why high quantum
efficiency is desirable: obviously, the number of extra qubits needed
is less, but also, it is desirable for error correction (although not
absolutely necessary) that the whole measurement be complete, and the
measurement outcome available, in a time much shorter (by a factor of
100, say) than the decoherence time.

Indeed, we think that there is a real prospect that such fast,
high-sensitivity measurements can be achieved for spins.  This is so
despite the fact that direct, magnetometric measurements of the spin
via its magnetic moment have, after many years of effort\cite{Rugar},
so far failed to achieve anything close to single-spin sensitivity; it
is for this reason that we are pessimistic about a
magnetic-force-microscope based quantum computer\cite{Berm}.  We feel
that a very promising general strategy involves first converting the
spin degree of freedom of the electron, which is very hard to measure,
into a charge degree of freedom, which is measurable at the
single-electron level by well-established electrometric techniques.
There are several possible ways that this could be achieved; we have
previously discussed\cite{LD,DMMM} a scheme in which a partially
transparent barrier (say, the bottom wall of the quantum dot) has a
barrier height which is spin-dependent, either because of a large
g-factor or an exchange splitting of the barrier.  Then, the measurement
would consist of applying a bias voltage to the dot so that the
electron is pressed up against this barrier; if the spin is up, say,
the barrier height is low, so the electron has a high chance of
tunneling and being detected by an electrometer circuit underneath;
while if the spin is down, the barrier height is high, tunneling does
not take place, and the electron is not detected.

In the donor-impurity scheme we can use a similar approach, which uses
some interesting physics of the donor impurity that was pointed out by
Kane\cite{Kane}.  The phosphorus impurity in Si has a stable singly
charged state (two electron state), but only if the relative spin of
the two electrons is a singlet.  If we wish to measure whether the
spins on two neighboring Ps are in a singlet or triplet (not identical
to a single spin measurement, but almost equally effective in
satisfying Criterion 5), a bias voltage is applied between the two Ps
so that the electron will tunnel from one P to another if a final
state is available (the singlet) but not otherwise.  Then, again,
electrometry can be used to determine whether the P has become charged
or not.

\subsection{Superconducting Qubits}

We will not review the superconducting proposals in detail, but we
will give a brief idea of how the five criteria are to be satisfied
and indicate to the reader where he can find more information on these
developments.  The proposals fall into two broad classes: ones in
which qubits involve the charge degree of freedom of the
superconductor, and ones where the qubits are embodied by the flux
states in a superconducting structure.

The charge models\cite{Averin,Makh} use as their basic structure a
pair of small superconducting islands separated by a Josephson
junction.  The isolation of the islands causes the Cooper-pair number
on the islands to be almost a good quantum number; a qubit can be
defined as one in which a Cooper pair is resident on either the left
or the right island.  Changing the voltages on the islands and,
possibly, the strength of the effective Josephson coupling, provide,
in principle, sufficient control to do one-bit gates, although it is
not clear whether the necessary parameters can be controlled with
sufficient precision, and whether ``1/f'' phenomena (random switching
of impurities near the device) can be adequately suppressed.  Two-bit
gate action is obtained by electrostatic coupling between
islands\cite{Averin} or coupling via the quantum states of an LC
circuit\cite{Makh}.  This coupling is not formally scalable to large
numbers of qubits, but it would be feasible for modest (10) numbers of
qubits. Readout would be by single-electron electrometry, which is
reasonably well understood.  The standard phenomenological model of
Josephson circuits suggest that the decoherence rates will be
low enough that gate operations can be performed.

The flux models come in a number of varieties\cite{Ioffe,Zagoskin};
the approach favored by the experimentalists\cite{Mooij} involves a
low-Tc (Al or Nb) SQUID circuit which, classically, has two
degenerate low energy states which differ slightly in their flux
configuration; from the quantum point of view, this is a double-well
potential problem with two nearly degenerate ground states, which
functions as the qubit.  The observation of single-qubit control is
then equivalent to the ``MQC'' phenomenon which has been sought in
these structures for many years.  Controlled inductive coupling
between SQUIDs is proposed for two-qubit gate action.  Again,
solutions for the five requirements are all proposed; it is possibly
worrisome that ``extrinsic'' effects resulting from the large-scale
motion of magnetic flux which might cause decoherence, e.g., coupling
to stray spins in the substrate, have apparently not been evaluated.

\subsection{Optical}

We would finally like to give a very brief mention of another
general line of attack that has some promise for solving some of the
problems contained in the five criteria: the use of solid-state
optical physics.  We have not studied these schemes in great detail,
but it is clear that they have the prospect of providing some of the
best solutions to the problems, for example, of single-quantum
measurement.  It is also clear that optical expertise is of great
value in the characterization of the quantum behavior of spins; all
the recent determinations of decoherence times of spins in
semiconductors\cite{Kikkawa} were performed using optical techniques.

So far, many of these proposals are not fully worked out; for the most
part, they conceptually follow the quantum-optics proposal involving
trapped atoms in a small optical cavity of Pellizzari {\em et
al.}\cite{Pel}.  For example, the proposal of Brun and Wang \cite{BW}
uses a very familiar object from quantum optics, the
whispering-gallery optical modes of a silica microsphere, to couple
quantum dots.  Other optical cavities with reasonably high Q are
available in solid state physics: in the proposal of Imamoglu and
coworkers\cite{Imam}, the microdisk optical cavity, a common structure
in laser research, is studied; in this structure the optical cavity is
formed by a pair of circular mirrors created by the deposition of
multiple layers of different III-V semiconductor materials (typically)
with different dielectric constants.  the mode volume is typically
also occupied by another III-V semiconducting material, but it can
also contain a collection of quantum dots.  In a manner related to
\cite{Pel} (and also to the more recent proposal \cite{SM}), the
quantized cavity modes provide a means of turning on and off two-bit
interactions between individual spins in the quantum dots (again, the
qubit of choice).  Single-qubit operations are accomplished by
near-resonant two-beam Raman transitions obtained by directing
(classical) light of the right frequency at particular quantum dots.
The decoherence properties are estimated to be marginal given the
current state of the art of these microcavities, but this technology
is expected to continue to improve.

This and other proposals are at a concrete enough stage that
experiments can begin to probe the ability to satisfy some of the five
criteria.  Many related approaches are possible, and we only mention
the electric-dipole transition qubits proposed by Sherwin {\em et
al.}\cite{Sher} (see also \cite{Sand}), the use of transitions in
optical hole burning materials\cite{Shah}, and the use of the
spectroscopy of excitons in quantum wells in III-V
materials\cite{SteGam}.  We don't want to predict where any of these
approaches will lead us, but they all have the possibility of future
success.

\section{Quantum Communication with Electrons}

The essential resources for quantum communication\cite{Bennett84} are
EPR (Einstein-Podolsky-Rosen) pairs\cite{Einstein}---pairwise
entangled qubits---the members of which are shared between
two parties (``Alice and Bob"). These parties are located at different
places and their goal for instance is to communicate with each other in an
absolutely secure way (which is not possible with classical means only).
The
prime example of an EPR pair considered here is the singlet state formed
by two electron spins. The intrinsic non-locality of these states gives
rise to striking phenomena such as violations of Bell inequalities and
has a number of possible applications in quantum information such as
quantum teleportation, quantum key distribution, and entanglement
purification.  The non-locality of EPR pairs has been
experimentally investigated for photons\cite{Aspect,Zeilinger}, but not
yet for {\it massive} particles such as electrons, let alone in a solid
state environment. This is so because it is difficult to first produce
and to then detect entanglement of electrons in a controlled way.  In the
following we review two scenarios we have recently
proposed\cite{LossSukhorukov,BLS} where the entanglement of electrons
(once produced e.g. as described in the previous sections) can be
detected in mesoscopic transport and noise measurements. One goal of the
following discussion is to show that there exists an interesting
connection between the field of quantum communication and the field of
transport theory in electronic nanostructures. Another goal is to show
that the investigation and concrete tests of fundamental phenomena such
as quantum non-locality for electrons are within experimental reach
within the not-so-distant future.  Such phenomena go beyond the standard
single-particle interference effects which have been well studied in
mesoscopic systems over the last decade or so.  Instead, they involve
genuine two-particle effects where, due to strong correlations leading to
entanglement, the quantum phases of two identical particles interfere
with each other in a constructive or destructive way. This two-particle
interference manifests itself in observable Aharonov-Bohm phase
oscillations in the electric transport current and in the non-equilibrium
current-current correlations.

\subsection{Probing Entanglement of Electrons in a Double Dot}

We consider a double-dot (DD) system  which contains two
metallic leads which are in equilibrium with associated reservoirs kept at
the
chemical potentials $\mu_{1,2}$.
Each lead is weakly coupled to {\it both} dots with
tunneling amplitudes $\Gamma$, and these leads are probes where the currents
$I_{1,2}$ are measured. Note that the DD system is put in parallel in
contrast to
the standard  situation where the coupled dots are put in series (i.e.
lead1-dot1-dot2-lead2). The quantum dots contain one (excess) electron each,
and are coupled to each other by the tunneling amplitude $t$, which
leads to a level splitting \cite{LD,Burkard}
$J=E_{\rm t}-E_{\rm s}\sim 4t^2/U$ in the DD,  with $U$ being the single-dot
Coulomb repulsion energy, and $E_{{\rm s}/{\rm t}}$ are the singlet/triplet
energies. We recall that for two electrons in the DD
(and for weak magnetic fields) the ground state is given by a spin
singlet.
For convenience we count the chemical potentials $\mu_i$ from $E_{\rm s}$.

The tunneling Hamiltonian \cite{Mahan} reads
$H=H_0+V$, where  $H_0=H_{\rm D}+H_1+H_2$ with $H_{\rm D}$ describing
the DD and $H_{1,2}$ the leads
(assumed to be Fermi liquids).
The tunneling between leads  and dots is described by
$V=V_1+V_2$,
where
\begin{equation}\label{perturbation}
V_n=\Gamma\sum_s\left[D^{\dag}_{n,s}c_{n,s}+c^{\dag}_{n,s}D_{n,s}\right],
\quad
D_{n,s}=e^{\pm i\varphi/4}d_{1,s}+e^{\mp i\varphi/4}d_{2,s} \enspace ,
\end{equation}
and where $c_{n,s}$ and $d_{n,s}$, $n=1,2$, annihilate
electrons with
spin $s$ in the $n$th lead and in the $n$th dot,  resp.
The Peierls phase $\varphi$ in the hopping
amplitude accounts for an Aharonov-Bohm (AB) or Berry phase (see below) in
the
presence of a magnetic field. The upper
sign belongs to lead 1 and the lower  to lead 2.
The average current
through the DD system is $I=\langle I_2\rangle$ with $I_n=
i e\Gamma\sum_s\left[D^{\dag}_{n,s}c_{n,s}-c^{\dag}_{n,s}D_{n,s}\right]$.

We consider now the Coulomb blockade (CB) regime where we can neglect
double (or higher) occupancy in each dot for all transitions including
virtual ones, i.e. we require $\mu_{1,2}<U$.  Further we assume that
$\mu_{1,2} >J, k_{\rm B}T $ to avoid resonances which might change the
DD state.  $\Gamma$ is assumed to be weak (i.e. $J> 2\pi\nu_{\rm t}
\Gamma^2$, where $\nu_{\rm t}$ is the tunneling density of states of
the leads) so that the state of the DD is not perturbed; this will
allow us to retain only the first non-vanishing contribution in
$\Gamma$ to $I$.  In analogy to the single-dot case
\cite{averinazarov,Konig} we refer to the above CB regime as the
cotunneling regime.  In leading order, the cotunneling current
involves the tunneling of one electron from the DD to, say, lead 1 and
of a second electron from lead 2 to the DD.  However, due to the weak
coupling $\Gamma$, the DD will have returned to its equilibrium state
{\it before} the next electron passes through it.  We focus on the
regime, $|\mu_1-\mu_2|>J$, where elastic and inelastic cotunneling
occurs, with singlet and triplet contributions being different.  In
this regime we can neglect the dynamics generated by $J$ compared to
the one generated by the bias,
and we finally obtain (for $k_BT<\mu_{1,2}$)
\begin{eqnarray}
&&I=e\pi\nu_{\rm t}^2\Gamma^4C(\varphi ){{\mu_1-\mu_2}\over {\mu_1\mu_2}}
\enspace ,\\ \label{current3}
&&C(\varphi )=\sum_{s,s'}\left[\langle
d^{\dag}_{1s^{\prime}}d_{1s}d^{\dag}_{1s}d_{1s^{\prime}}\rangle +\cos
\varphi\langle d^{\dag}_{1s^{\prime}}d_{1s}d^{\dag}_{2s}d_{2s^{\prime}}
\rangle \right] \enspace .
\label{factor1}
\end{eqnarray}
Eq. (\ref{current3})
shows that the cotunneling
current depends on the properties of the ground state of the DD
through the coherence factor $C(\varphi )$ given in  (\ref{factor1}).
The first term in $C$ is the contribution
from the topologically trivial tunneling path which runs from
lead 1 through, say, dot 1 to lead 2 and the same path back. The second
term (phase-coherent part) in $C$ is the ground state amplitude of  the
exchange
of electron
1 with electron 2 via leads 1 and 2 such that a closed loop is formed
enclosing an area $A$.
Thus, in the presence of a magnetic
field $B$, an AB phase factor  $\varphi=ABe/h$ is acquired.

Next, we evaluate $C(\varphi)$ explicitly
in the singlet-triplet basis.
Note that only the singlet
$|S\rangle$ and the triplet $|T_0\rangle$
are entangled EPR pairs while
the remaining triplets
$|T_{+}\rangle=|\!\uparrow\uparrow\rangle$, and
$|T_{-}\rangle =|\!\downarrow\downarrow\rangle$
are not (they factorize).
Assuming that the DD is in one of these states we obtain
\begin{equation}\label{factor2}
C(\varphi )=
\left\{\begin{array}{ll}
2-\cos \varphi \, , \quad & \mbox{for the singlet,}\\
2+\cos \varphi \, , \quad & \mbox{for all triplets.}\enspace
\end{array}\right.
\end{equation}
Thus, we see that the singlet and the triplets contribute with {\it
opposite sign to the phase-coherent part of the current}.  One has to
distinguish, however, carefully the entangled from the non-entangled
states.  The phase-coherent part of the entangled states is a genuine
{\it two-particle} interference effect, while the one of the product
states cannot be distinguished from a phase-coherent {\it
single-particle} interference effect.  Indeed, this follows from the
observation that the phase-coherent part in $C$ factorizes for the
product states $T_{\pm}$ while it does not do so for the entangled
states $S, T_0$.  Also, for states such as
$|\!\uparrow\downarrow\rangle$ the coherent part of $C$ vanishes,
showing that two different (and fixed) spin states cannot lead to a
phase-coherent contribution since we {\it know} which electron goes in
which part of the loop.  Finally we note that due to the AB phase the
role of the singlet and triplets can be interchanged, which is to say
that we can continually transmutate the statistics of the entangled
pairs $S,T_0$ from fermionic to bosonic (like in anyons): the
symmetric orbital wave function of the singlet $S$ goes into an
antisymmetric one at half a flux quantum, and vice versa for the
triplet $T_0$.

The amplitude of the AB oscillations is a direct measure of the phase
coherence of the entanglement, while the period via the enclosed area
$A=h/eB_0$ gives a direct measure of the non-locality of the EPR
pairs, with $B_0$ being the field at which $\varphi=1$.
Thus, the measurement of the AB amplitude will provide us with
an entanglement dephasing length, which tells us how far we can
spatially separate two electrons from each other in a conductor
(in the presence of many other electrons, spin-orbit interaction,
spin-impurities, etc.) before the entanglement in the total spin state
is lost. No doubt, it would be highly desirable to obtain
experimental information about this length scale since this will
allow us to
assess if and under which conditions quantum communication (which
makes essential use of separated EPR pairs)
will be possible in mesoscopic structures.

The triplets
themselves can be further distinguished by applying a directionally
inhomogeneous magnetic field (around the loop) producing a Berry phase
$\Phi^{\rm B}$ \cite{LossBerry}, which is positive (negative) for the
triplet $m=1 (-1)$, while it vanishes for the EPR pairs $S,
T_0$. Thus, we will eventually see beating in the AB oscillations due
to the positive (negative) shift of the AB phase $\Phi$ by the Berry
phase, $\varphi=\Phi \pm \Phi^{\rm B}$. We finally note that the
closed AB-loop can actually be made as large as the dephasing length
by using wave guides forming a loop with leads attached to it. Thus, a
moderately weak field can be applied to produce the AB oscillations
with negligible effect of the orbital state of the DD.

We discuss now the spectral density (noise) of the current
cross-correlations, $S(\omega)=\int dte^{i\omega t}Re\langle\delta
I_2(t)\delta I_1(0)\rangle$.  Under the same assumptions as before
(cotunneling regime), we obtain for the zero-frequency noise its
Poissonian value, i.e.  $S(0)=-e|I|$. This shows that the Fano factor
(noise-to-current ratio) is universal and the current and its
cross-correlations contain the same information. For finite
frequencies in the regime $|\mu_1-\mu_2|>J$ and at $T=0$, we find
$S(\omega)=(e\pi\nu_t\Gamma^2)^2 {C(\varphi )}
\left[X_{\omega}+X^{*}_{-\omega}\right] $, where
\begin{eqnarray}\label{noise2}
& &ImX_{\omega}=\left[\theta (\mu_1-\omega
)-\theta (\mu_2-\omega )\right]/2\omega,  \\
& & ReX_{\omega}={1\over {2\pi\omega}}sign(\mu_1-\mu_2+\omega
)\ln|{{(\mu_1+\omega )(\mu_2-\omega )}\over {\mu_1\mu_2}}|
\nonumber \\
& & -{1\over {2\pi \omega }}\left[\theta (\omega -\mu_1)\ln
|{{\mu_2-\omega }\over {\mu_2}}|+
\theta (\omega -\mu_2)\ln
|{{\mu_1-\omega}\over {\mu_1}}|\right].
\end{eqnarray}
The noise again depends on the phase-coherence factor $C$ with the
same properties as discussed before.
Here, Re$ S(\omega)$ is even in $\omega$, while Im$ S(\omega)$
is non--zero (for finite frequencies) and odd,
in contrast to single-barrier junctions, where Im$ S(\omega)$
vanishes, since $\delta I_1=-\delta I_2$ for all times.
At small bias
$\Delta\mu=\mu_1-\mu_2\ll\mu =\left(\mu_1+\mu_2\right)/2$, the odd part,
Im$ S(\omega)$, given in (\ref{noise2})
exhibits  two narrow peaks at $\omega =\pm\mu$, which  lead to
slowly decaying oscillations in time,
$S_{odd}(t)=
\pi\nu_t^2\Gamma^4C(\varphi )\sin(\Delta\mu t/2)\sin (\mu t)/\mu t$.
These oscillations can be interpreted as a temporary charge-imbalance
on the DD during an uncertainty time $\sim \mu^{-1}$, which results
from the cotunneling of electrons and an associated time shift
(induced by a finite $\omega$) between incoming and outgoing currents.

There are a few obvious generalizations to the material
presented so far: (1) multi-dot and multiterminal set-ups
which implement  n-particle entanglement, a prime example being the
3-particle
entangled GHZ states {\it etc.};
(2) variations of the geometries such as the phase-coherent transport
from additional ``feeding leads" into dots 1 and 2. Such a set-up
corresponds
topologically to a  scattering experiment in which we can arrange for
scattering of unentangled  electrons (as considered previously in
noise studies\cite{noise}) but now also of entangled ones. In the latter
case we get a non-trivial Fano factor\cite{BLS} due to
antibunching (triplets) and bunching (singlet) effects in the
noise\cite{MMM}: see below. (3) We can replace leads 1 and 2
each by
quantum dots which are connected to the double-dot by spin-selective
tunneling
devices\cite{Prinz} (such spin-filters would allow us to measure spin  via
charge\cite{LD}).  Such or similar set-ups would be needed to measure
all spin
correlations contained  in the electronic EPR pairs and thus to test Bell
inequalities for electrons in a solid state environment.

\subsection{Noise of Entangled Electrons: Bunching and Antibunching}

In this section we discuss a related but alternative scenario in which
entanglement of electrons can be measured through a
bunching and antibunching behavior in the noise of
conductors\cite{MMM,BLS}.  The basic idea is rather simple and
well known from the scattering theory of two identical
particles\cite{Feynman,Ballentine}. In the center-of-mass system the
differential
scattering cross-section can be expressed in terms of the  scattering
amplitude $f(\theta)$  and scattering  angle $\theta$ \cite{Ballentine},
\begin{equation}
\sigma(\theta)=|f(\theta) \pm f(\pi-\theta)|^2=|f(\theta)|^2+|
f(\pi-\theta)|^2 \pm 2 Re f^*(\theta)f(\pi-\theta).
\end{equation}
The first two terms in the second equation are the ``classical''
contributions
which are obtained if the particles were distinguishable, whereas the third
term
results from the indistinguishability which gives rise to constructive
(destructive) {\it two-particle interference effects}. Here the plus sign
applies
for spin-1/2 particles in the singlet state (described by a symmetric
orbital wave
function), while the minus sign applies for their triplet states (described
by an
antisymmetric orbital wave function). The
very same two-particle interference mechanism which is responsible for the
enhancement/reduction of the  scattering cross section $\sigma$ near
$\theta=\pi/2$
leads to a bunching/antibunching behavior in the statistics\cite{Loudon}.

We have previously described in detail how two electron spins can be
deterministically entangled by weakly coupling two nearby quantum
dots, each of which contains one single (excess)
electron\cite{LD,Burkard}. The recently investigated coupling between
electrons which are trapped by surface acoustic waves on a
semiconductor surface\cite{Barnes} might provide another possibility
of producing EPR pairs in a solid-state environment. Generalizing the
above two-particle scattering experiment to a mesoscopic system, we
have discussed an experimental set-up by which the entanglement of
electrons (moving in the presence of a Fermi sea) can be detected in
measurements of the current correlations (noise)\cite{MMM,BLS}.  For
this purpose we employ a beam splitter which has the property that
electrons fed into its two incoming leads have a finite amplitude to
be interchanged (without mutual interaction) before they leave through
the two outgoing leads.  In our case, the electrons are entangled
before they enter the beam splitter.  The quantity of interest is then
the noise measured in the outgoing leads of the beam splitter.  It is
well-known that particles with symmetric wave functions show bunching
behavior\cite{noise} in the noise, whereas particles with
antisymmetric wave functions show antibunching behavior. The latter
situation is the one considered recently for electrons in the normal
state of mesoscopic transport systems both in
theory\cite{Buettiker1,Martin} and in
experiments\cite{Stanford,MartinSC}.  However, since the noise is
produced by the charge degrees of freedom we can expect\cite{MMM} that
in the absence of spin scattering processes the noise is sensitive to
the symmetry (singlet or triplet) of only the {\it orbital part} of
the wave function.  We have verified this expectation explicitly, by
extending the standard scattering matrix approach for transport in
mesoscopic systems\cite{Buettiker1} to a situation with
entanglement\cite{BLS}.

The electron current operator in lead $\alpha$ of a multiterminal
conductor is
\begin{equation}
  I_{\alpha}(t) =
\frac{e}{h\nu}\sum_{E,E',\sigma}\left[
a_{\alpha\sigma}^\dagger(E)a_{\alpha\sigma}(E')-b_{\alpha\sigma}^\dagger(E)
b_{\alpha\sigma}(E')
\right]e^{i(E-E')t/\hbar},\label{current_def}
\end{equation}
where $a^\dagger_{\alpha \sigma} (E)$ creates an incoming electron in lead
$\alpha$ with spin $\sigma$ and energy $E$, and the operators
$b_{\alpha\sigma}$ for the outgoing electrons
are related to the operators $a_\alpha$ for the incident electrons via
$s_{\alpha\beta}$, the (spin- and energy-independent) scattering matrix,
$b_{\alpha\sigma}(E)=\sum_{\beta} s_{\alpha\beta}a_{\beta\sigma}(E)$.
Note that since we are dealing with discrete energy states here, we
normalize the operators $a_\alpha(E)$ such that
$\left\{a_{\alpha\sigma}(E),a_{\beta\sigma'}(E')^\dagger\right\}=
\delta_{\sigma\sigma'}\delta_{\alpha\beta}\delta_{E,E'}/\nu$,
where the Kronecker symbol $\delta_{E,E'}$ equals $1$ if $E=E'$ and $0$
otherwise, and $\nu$ stands for the density of states in the leads.
We also assume that each lead consists of only a single quantum channel;
the generalization to leads with several channels is straightforward but is
not needed here.

We evaluate the spectral density for the current fluctuations
$\delta I_{\alpha}=I_{\alpha}-\langle I_{\alpha}\rangle$
between the leads $\alpha$ and $\beta$,
\begin{equation}
  \label{cross1}
  S_{\alpha\beta}({\omega})
  = \lim_{T\rightarrow\infty}
  \frac{h\nu}{T}\int_0^T\!\!\!dt\,\,e^{i\omega t}
  \langle\Psi|\delta I_{\alpha}(t)\delta I_{\beta}(0)|\Psi\rangle,
\end{equation}
for the entangled incident state
\begin{equation}
  \label{entangled_state}
  |\Psi\rangle = |\pm\rangle
  = \frac{1}{\sqrt{2}}\left( a_{2\downarrow}^\dagger(\epsilon_2)
a_{1\uparrow}^\dagger(\epsilon_1)
    \pm a_{2\uparrow}^\dagger(\epsilon_2)
a_{1\downarrow}^\dagger(\epsilon_1)\right) |0\rangle .
\end{equation}
The state $|-\rangle$ is the spin singlet, $|S\rangle$,
while $|+\rangle$ denotes one of the spin triplets
$|T_{0,\pm}\rangle$; in the following we will present a calculation of
the noise for $|+\rangle=|T_0\rangle$, i.e. the triplet with $m_z=0$.
Evaluating the matrix elements we obtain the current correlation
between the leads $\alpha$ and $\beta$,
\begin{equation}
  S_{\alpha\beta}(0)
   = \frac{e^2}{h\nu}\left[\sum_{\gamma\delta}\!{}^{'}
    A_{\gamma\delta}^{\alpha}A_{\delta\gamma}^{\beta}
   \mp \delta_{\epsilon_1,\epsilon_2}
    \left(A_{12}^{\alpha}A_{21}^{\beta} +A_{21}^{\alpha}A_{12}^{\beta})
 \right)\right],\label{cross5}
\end{equation}
where $A_{\beta\gamma}^{\alpha} = \delta_{\alpha\beta}\delta_{\alpha\gamma}
-s_{\alpha\beta}^{*} s_{\alpha\gamma}$, and
$\sum_{\gamma\delta}^{\prime}$ denotes the sum over $\gamma=1,2$ and
all $\delta\neq\gamma$, and where again the upper (lower) sign refers to
triplets (singlets).

We apply these formulas now to our scattering set-up involving a beam
splitter
with four attached leads (leads $1$ and $2$ incoming, leads $3$ and $4$
outgoing)
described by the single-particle scattering matrix elements,
$s_{31}=s_{42}=r$,
and $s_{41}=s_{32}=t$, where $r$ and $t$ denote the reflection and
transmission
amplitudes at the beam splitter, respectively.
We assume that there is no backscattering,
$s_{12}=s_{34}=s_{\alpha\alpha}=0$.
The unitarity of the s-matrix implies $|r|^2+|t|^2=1$.
The final result for the noise correlations for the
incident state $|\pm\rangle$ is then\cite{footnote1},
\begin{eqnarray}
S_{33}(0)=S_{44}(0)=2\frac{e^2}{h\nu}T\left(1-T\right)
  \left(1\mp \delta_{\epsilon_1,\epsilon_2}\right),\label{noise_f1}\\
S_{34}(0)=2\frac{e^2}{h\nu}Re\left[{r^*}^2t^2\right]
  \left(1\mp \delta_{\epsilon_1,\epsilon_2}\right),\label{noise_f2}
\end{eqnarray}
where $T=|t|^2$ is the probability for transmission through the
beam splitter.
The calculation for the remaining two triplet states
$|+\rangle=|T_\pm\rangle=|\uparrow\uparrow\rangle
,|\downarrow\downarrow\rangle$
yields the same results Eqs.~(\ref{noise_f1}) and
(\ref{noise_f2}) (upper sign).
For the average current in lead $\alpha$ we obtain
$\left|\langle I_\alpha\rangle\right| = e/h\nu$,
with no difference
between singlets and triplets.
Then, the Fano factor
$F = S_{\alpha\alpha}(0) /\left|\langle I_\alpha\rangle\right|$
takes the form
\begin{equation} \label{fano}
  F =  2eT(1-T)\left(1\mp \delta_{\epsilon_1,\epsilon_2}\right),
\end{equation}
and correspondingly for the cross correlations.
This result implies that if two electrons
with the same energies, $\epsilon_1=\epsilon_2$, in the singlet
state $|s\rangle = |-\rangle$ are injected into lead $1$ and lead $2$,
resp.,
then the zero frequency noise is {\it enhanced} by a factor of two,
$F=4eT(1-T)$, compared to the shot noise of uncorrelated particles,
$F=2eT(1-T)$. This enhancement of noise is
due to {\it bunching} of electrons in the outgoing leads, caused by the
symmetric orbital wavefunction of the spin singlet $|s\rangle$.
On the other hand, the triplet states $|+\rangle = |T_{0,\pm}\rangle$
exhibit an {\it antibunching} effect, leading to a complete
suppression of the zero-frequency noise, $S_{\alpha\alpha}(0)=0$.
The noise enhancement for the singlet $|S\rangle$ is a
unique signature for entanglement (there exists no unentangled state with
the same symmetry), therefore entanglement can be observed by
measuring the noise power of a mesoscopic conductor.
The triplets can be further distinguished from each other
if we can measure the
spin of the two electrons in the outgoing leads, or if we insert
spin-selective tunneling devices\cite{Prinz} into leads 3,4
which would filter a certain spin polarization.

Note that above results remain unchanged if we consider
states $|\pm \rangle$ which are created above a Fermi sea. We have shown
elsewhere\cite{MMM} that the entanglement of two electrons
propagating
in a Fermi sea gets reduced by the quasiparticle weight $z_F$ (for each
lead one factor)
due to the presence of interacting electrons.
In the metallic regime $z_F$ assumes typically some finite
value\cite{qpweight},
and thus as long as spin scattering processes are small the above
description
for non-interacting electrons remains valid.

\section{Conclusion}

We hope that workers in mesoscopic physics will  find this brief
survey of recent theoretical developments in quantum computation
stimulating.  As we continue to learn about quantum computing and
quantum communication, we see more and more connections with
present-day experimental physics.  Quantum computing is not just a
mathematical abstraction, it changes our outlook on a variety of
fundamental issues in mesoscopics: on the desirability of having long
coherence times in mesoscopic structures, on the role of precise
time-dependent control of these structures for manipulating the
interaction of electron states, on the need to develop high quantum
efficiency measurements for spin and other single-quantum properties.
Quantum computing and communication clearly have a fascinating role to
play in some far-future technologies; we hope that we have illustrated
how they can also play a role in the direction of fundamental physics
research today.

\section*{Acknowledgments}
We would like to thank K. Ensslin for kindly providing us with
essential material parameters, and A. Chiolero for advising us on our
numerical method for the g-factor calculation.  DPD is grateful for
funding under grant ARO DAAG55-98-C-0041.  GB, DL, and EVS acknowledge
the funding from the Swiss National Science Foundation.


\begin{thebibliography}{}
\bibitem{footnote1}
For finite frequencies, we obtain the noise power
$S_{\alpha\alpha}(\omega) = (2e^2/h\nu)[(1-\delta_{\omega,0})
+2T(1-T)(\delta_{\omega,0}\mp\delta_{\omega,\epsilon_1-\epsilon_2})]$.
\bibitem{qpweight} For instance, in metals such as bulk Cu the
quasiparticle weight becomes, within the RPA approximation,
$z_F=0.77$ \protect\cite{Rice}, while for a GaAs 2DEG we find (also
within RPA) $z_F=1-r_s(1/2+1/\pi)=0.66$ for the GaAs interaction
parameter $r_s=0.61$ (the details of the calculation will be given
elsewhere).
\end{thebibliography}
\end{document}